\documentclass[twocolumn,PRB,amsmath,amssymb]{revtex4-1}
\usepackage{amssymb}
\usepackage{amsmath}
\usepackage{graphicx}
\usepackage{epstopdf}
\usepackage{bm}%
\usepackage{gensymb}
\usepackage{soul}
\usepackage{sidecap}
\usepackage[normalem]{ulem} 
\usepackage{epsfig}
\usepackage{graphicx}
\usepackage{amsfonts,amssymb}
\usepackage{amsmath}
\usepackage{color}
\usepackage {times}
\usepackage{epstopdf}

\newcommand{\bl}{\begin{aligned}}
\newcommand{\el}{\end{aligned}}

\newcommand{\be}{\begin{equation}}
\newcommand{\ee}{\end{equation}}   
\newcommand{\bea}{\begin{eqnarray}}
\newcommand{\eea}{\end{eqnarray}}
\newcommand{\ba}{\begin{array}}
\newcommand{\ea}{\end{array}}

\newcommand{\q}{{\bf q}}
\renewcommand{\k}{{\bf k}}
\newcommand{\Q}{{\bf Q}}

\graphicspath{{Figs/}}

\begin{document}

\title{Unidirectional charge correlations in hole-doped cuprates}

\date{\today}
\author{Dheeraj Kumar Singh}
\author{Yunkyu Bang}
\email{dheeraj.kumar@thapar.edu }
\email{ykbang@apctp.org}
\affiliation{School of Physics and Materials Science, Thapar Institute of Engineering and Technology, Patiala-147004, Punjab, India}
\affiliation{Department of Physics, POSTECH, Pohang 790-784, Korea}
\affiliation{Asia Pacific Center for Theoretical Physics, Pohang, Gyeongbuk 790-784, Korea}

%
\begin{abstract}
We examine charge correlations and instabilities in the pseudogap phase of high-$T_c$ cuprates modeled by $d$-density wave ordering. The latter has a gap symmetry similar to the one observed in the $d$-wave superconductor. We use $t$-$J$ model to describe the charge correlations in the presence of electron-phonon interaction. Our finding suggest that the charge instability in the normal state is dominating at an  incommensurate wavevector along (1, 1) instead of (1, 0) for a realistic interaction parameter. The dominance at the diagonal wavevector is further enhanced if the coupling between electron and bond-buckling $B_{1g}$ phonon is incorporated. On the other hand, a dominating charge-density correlation develops along (1, 0) at an incommensurate wavevector in the $d$-density wave ordered state, which shows a qualitative agreement with the experiments. The correlation becomes robust only in the presence of $B_{1g}$ phonon.
\end{abstract}
\maketitle
\section{Introduction} 
After the discovery of unconventional superconductivity in the hole-doped cuprates in late 90s~\cite{bednorz}, an incessant advancements in the experimental techniques have unveiled plethora of properties exhibited by the so-called \textit{pseudogap} phase. The pseudogap phase is observed in a wide region of hole doping located above the domes associated with long-range magnetic order and $d$-wave superconductivity in the temperature-vs-doping phase diagram of high-$T_c$ cuprates~\cite{mueller,robinson}. Thermal, transport and spectral properties~\cite{timusk} of this phase have often been associated with different symmetry-breaking phenomena including nematicity~\cite{sato}, charge-density wave (CDW)~\cite{frano}, pair-density wave (PDW)~\cite{du}, loss of inversion symmetry~\cite{zhao} etc. Possible presence of multiple symmetries breaking leads to diverging scenarios, which complicates the longstanding issue of actual origin of the pseudogap further~\cite{renner,ding,loeser,norman,yoshida,kanigel}. 

Evidences for the onset of incommensurate charge correlations and ordering in the portion of the region occupied by pseudogap phase just above the superconducting dome have been obtained through several experiments including neutron-scatterings~\cite{tranquada}, resonant-inelastic x-ray scattering (RIXS)~\cite{ghiringhelli,comin}, x-ray diffraction~\cite{chang,canosa}, scanning-tunneling microscopy~\cite{kohsaka,wise} etc. The charge correlations were speculated to arise from the instability associated with  Fermi-surface topology specific to the pseudogap phase. However, several experiments suggest otherwise because these charge correlations are also observed even beyond the critical doping $x_c \sim 0.19$ where the pseudogap features are absent~\cite{miao} or in the electron-doped cuprates~\cite{jang}. While the CDW correlations have been observed in majority of the cuprates including YBa$_2$Cu$_3$O$_{6+x}$~\cite{ghiringhelli}, there are evidences of coexisting charge-spin superstructures in La$_2$CuO$_ 4$~\cite{tranquada}. 
These strong incommensurate charge fluctuations may develop into a long-range order in the presence of external magnetic field~\cite{chang1}, uniaxial~\cite{kim} and epitaxial strain~\cite{bluschke}. However, some of the experiments indicate their existence even without any external perturbation~\cite{hucker}.

All the incommensurate wavevectors associated with the CDW correlations are unidirectional and no signatures of bidirectional wavevectors have been obtained yet. The CDW  wavevector ${\bf q}^* \sim (0.6\pi, 0)$~\cite{ghiringhelli,comin,neto} is nearly same for various cuprates except for the Hg- and La-based~\cite{tabis,tranquada}. For the latter case, ${ q^*_x} \sim 0.55 \pi$ and $ 0.45\pi$, respectively, which is  slightly on lower side. It is further small for electron-doped cuprates~\cite{neto1} and may also show dependence on both temperature and doping~\cite{miao1}. 

The origin of strong charge fluctuations or long-range charge order  with uni- instead of bi-directional modulation vectors have often been associated with different factors. Recent works suggest that the PDW state with wave vector ${\bf q}^*$ may have the potential to induce CDW order with a wavevector $2{\bf q}^*$~\cite{norman1,agterberg}. One of the other possible candidates is the long-range repulsive Coulomb interaction considered in the N\'{e}el ordered state within the  three-orbital model~\cite{atkinson}. Interestingly, the Coulomb repulsion between two neighboring oxygen $p$-orbital electrons and between $s$- and $p$- orbital electrons were shown to play a very important role instead of Cu $d$-orbital electrons. However, the interaction parameter required to generate the instability was too large and comparable to the largest interaction parameter, i.e., the on-site repulsive Coulomb interaction for the relatively localized $d$ orbitals. Secondly, the N\'{e}el order is known to be fragile with respect to hole doping and associated ordered magnetic moments are inconsistent with the experiments. Moreover, the charge correlations set at the incommensurate wavevector for the intra-pocket scattering was not compared~\cite{atkinson}.   

Another potential clue about the origin  of CDW order is obtained from Raman-scattering spectroscopy which shows the  softening of dispersionless bond-buckling $B_{1g}$ phonon modes~\cite{cuk,forgan}. $B_{1g}$ phonon modes involve out-of-phase motion of different pairs of oxygen atoms lying on the opposite sides of the square lattice formed by the Cu atoms. The softening though small near the transition from the state with CDW correlations to $d$-wave superconducting state ($d$SC) gets enhanced further in the $d$SC state. 

In a recent work, it has been claimed that the charge susceptibility can be peaked at the incommensurate wavevectors observed in the experiment because of the coupling between the electron and $B_{1g}$ phonons. However, the Fermi surface for the pseudogap was modeled by adopting a direction dependent functional form of quasiparticle weight which decreases as one moves away from the nodal point along the normal-state Fermi surface~\cite{banerjee}. This is, however, in contrast with the results from the angle-resolved photoemssion spectroscopy (ARPES), because while the quasiparticle weight gets increasingly suppressed towards the antinodal point,  the quasiparticle peak does exist but slightly away from the Fermi level. Secondly, the concentration of quasiparticle weight in the vicinity of nodal points is expected to lead the charge correlation to develop near $(\pi, \pi)$ and $(\pi, 0)$ instead. The correlation at the former wavevector may get suppressed by the $B_{1g}$ phonon but not at the latter wavevector. Therefore, despite various efforts, it is not clear enough as to how the unidirectional charge correlations develop in the hole-doped cuprates at a wavevector close to $\sim (0.5\pi, 0)$.  

In this paper, we examine the onset of charge correlation in the normal as well as in the pseudogap phase. We model the pseudogap gap phase by $d_{x^2-y^2}$-density wave (DDW) ordering~\cite{sudip} with a gapstructure identical to the $d$wave superconductivity, which is one of the several leading candidates proposed to explain the highly unusual behavior of the pseudogap phase including the spectral properties. Unlike the unrealistic long-range magnetic order and associated large magnetic moments in the hole-doped cuprates considered in earlier work, the DDW state involves a very weak staggered magnetic moment arising due to circulating bond currents in a checkerboard pattern while the time-reversal symmetry is broken. This results in gapping out of the Fermi surface near anti-nodal points as observed in the experiments and small hole pockets in the vicinity of nodal points. Our investigation highlights the important role of both Coulomb interaction and electron-phonon coupling in setting up the unidirectional charge correlations resulting from the inter-pocket nesting which finds a stiff competition with the bidirectional charge correlation arising out from the intra-pocket nesting.  

\section{model and method} 
In order to investigate the charge-ordering instability in the DDW state, we consider the following phenomenological model
\bea
H_{D}(\k) &=& \sum_{\k,\sigma} (\varepsilon_{\k} -\mu)  d^{\dagger}_{\k \sigma}d_{\k \sigma} + \sum_{\k^{\prime},\sigma} i W_k d^{\dagger}_{\k+\Q  \sigma} d_{\k \sigma} + h.c. \nonumber\\ &=& \sum_{(\k),\sigma} \psi^{\dagger}_{\k \sigma} \mathcal{H}(\k) \psi_{\k \sigma}, 
\eea 
where $\varepsilon_{\k} = -2t_1 (\cos k_x + \cos k_y) + 4t_2 (\cos k_x \cos k_y) + -2t_3 (\cos 2k_x + \cos 2k_y)$ and $\mu$ is chemical potential. $t_1$, $t_2$, and $t_3$ are the nearest, next-nearest, and next-next nearest neighbor hopping parameters, respectively. In the calculation, $t_2 = 2t_1/5$ and $t_3 = t_1/8$. $\Q = (\pi, \pi)$ is the DDW state ordering wavevector. The DDW gap $W_k = W_0 f_{\k} = i \sum_{\k^{\prime}} V_{\k \k^{\prime}} \langle d^{\dagger}_{\k^{\prime}+\Q \sigma} d^{}_{\k^{\prime}\sigma} \rangle$, where $f_{\k} = \cos k_x - \cos k_y$ and $V_{\k \k^{\prime}}  = V_0 f_{\k} f_{\k^{\prime}}$.
$\psi^{\dagger}_{\k,\sigma} = (d^{\dagger}_{\k \sigma}, d^{}_{\k + \Q  \sigma})$ and 
\be
\mathcal{H}(\k) = \begin{pmatrix}
         \varepsilon_{\k} & iW_{\k+\Q} \\
         iW_{\k} &  \varepsilon_{\k+\Q} 
        \end{pmatrix} 
        =\begin{pmatrix}
         \varepsilon_{\k} & -iW_{\k} \\
         iW_{\k} &  \varepsilon_{\k + \Q} 
        \end{pmatrix} .
\ee
Diagonalization of $\mathcal{H}(\k)$ yields the eigenvalues
\be
E^{\alpha,\beta} = \varepsilon^+_{\k}  \pm \sqrt{(\varepsilon^-_{\k})^2+W^2_{\k}}  
\ee
where
\be
\varepsilon^{\pm}_{\k} = \frac{\varepsilon_{\k} \pm \varepsilon_{\k+\Q}}{2}
\ee
To investigate the charge correlation, we consider the charge susceptibility defined as follows:
\begin{equation}
\chi^{}(\q,i\Omega_n)= \int^{\beta}_0{d\zeta e^{i \Omega_{n}\zeta}\langle T_\zeta [{\cal \rho}_\q(\zeta) {\cal \rho}_{-\q}(0)]\rangle}.
\end{equation}
Here, $\langle...\rangle$ denotes thermal average, $T_\zeta$ imaginary time ordering, and $\Omega_n$ are the Bosonic
Matsubara frequencies. ${\cal \rho}_{\bf q}$ is 
obtained as the Fourier transformation of 
$\mathcal{\rho}_{{\bf i}} = \sum_{\sigma} n_{\bf i \rho} $. 

The static charge susceptibility in DDW state takes a form of 2 $\times$ 2 matrix 
\be
\hat{\chi}(\q) = \begin{pmatrix} 
                  \chi(\q,\q) & \chi(\q,\q+\Q) \\ 
                  \chi(\q+\Q,\q) & \chi(\q+\Q,\q+\Q)
                 \end{pmatrix}
\ee
with 
\be
\chi(\q,\q) = \sum_{\k} \bigl( c^{+}(\k,\q) \chi^{\alpha \alpha}_0 (\q) +   c^{-}(\k,\q) \chi^{\alpha \beta}_0(\q)  \bigl),
\ee
where 
\be
\chi^{\mu \nu}_0 (\q) = \sum_{\k} \chi^{\mu \nu}_0 (\k; \q)  = \sum_{\k} \frac{f(E^{\mu}_{\k+\q}) - f(E^{\nu}_{\k})}{ E^{\mu}_{\k} - E^{\nu}_{\k+\q} + i\eta }
\ee
and
\bea
      c^{\pm}(\k,\q) = \frac{1}{2} \biggl(1 \pm \frac{\varepsilon^-_{\k} \varepsilon^-_{\k+\q}+W_{\k}W_{\k+\q}}{(E^{\alpha}_{\k}-E^{\beta}_{\k}) (E^{\alpha}_{\k+\q}-E^{\beta}_{\k+q})}\biggl).
\eea       
The other diagonal term is 
\bea
\chi(\q+\Q,\q+\Q) &=&  \sum_{\k} \bigl( c^{-}(\k,\q) \chi^{\alpha \alpha}_0 (\q) \nonumber\\ &+&   c^{+}(\k,\q) \chi^{\alpha \beta}_0(\q)  \bigl).
\eea
The off-diagonal terms of the static-charge susceptibility is 
\bea
\chi(\q+\Q,\q,i\omega_n) & = & -\chi(\q+\Q,\q,i\omega_n) \nonumber\\ & = & \frac{1}{2} \sum_{\k}  c(\k,\q) \bigl( \chi^{\alpha \alpha}_0 (\q)  +  \chi^{\beta \beta}_0 (\q) \nonumber\\  &-& \chi^{\alpha \beta}_0 (\q) - \chi^{\beta \alpha}_0 (\q) \bigl),
\eea
where
\bea
      c^{}(\k,\q) = - \frac{i}{2} \frac{\varepsilon^-_{\k}W_{\k+\q} -\varepsilon^-_{\k+\q}W_{\k}}{(E^{\alpha}_{\k}-E^{\beta}_{\k}) (E^{\alpha}_{\k+\q}-E^{\beta}_{\k+q})} 
\eea       

The charge susceptibility at the level of random-phase approximation is given by 
\be
\hat{\chi}_{R}(\q) = (\hat{I}-\sum_{\k}\hat{C}\hat{\chi}(\k;\q))^{-1}\hat{\chi}(\q)
\ee
where $\hat{I}$ is a 2$\times$2 identity matrix and contribution to the interaction matrix $\hat{C}$ arises from the attractive density-density interaction term of $t$-$J$ model 
\be 
 H_{t-J} = J\sum_{\langle i,j \rangle} \left( {\bf S}_{i} \cdot {\bf S}_{j} - \frac{n_i n_j}{4}\right),
\ee
where $J \sim 4t_1^2/U$ with $U$ being the on site Coulomb interaction between the electrons of opposite spins in of an orbital. It may be noted that the first term doesn't contribute to the charge susceptibility.  Another contribution to $\hat{C}$ comes from the electron-phonon coupling given by 
\be
H_{e-ph}(\k) = \sum_{\k,\sigma} g(\k,\q) d^{\dagger}_{\k+\q \sigma} d^{}_{\k \sigma} (a^{\dagger}_{-\q} + a^{}_{\q}).  
\ee
Here, $g(\k,\q)$ is the electron-phonon coupling for the bond-buckling $B_{1g}$ phonon and $a^{}_{\q}$ is the phonon annihilation operator. $g(\k,\q)$ is given by~\cite{devereaux,devereaux1,devereaux2}
\bea
g(\k,\q) &=& 2eE_z \sqrt{\frac{\hbar}{2M_oN_{\q}\omega_{B_{1g}}}} 
(\phi^*_x(\k) \phi^*_x(\k-\q) \cos q_x/2 \nonumber\\ &-& \phi^*_y(\k) \phi^*_y(\k-\q) \cos q_y/2),
\eea
where 
$M_o$ is the mass of oxygen atom, $\omega_{B_{1g}} ~ 40$meV is energy of $B_{1g}$ is phonon. It is useful to define dimensionless electron-phonon coupling $\tilde{g}(\k,\q) =  g(\k,\q)/\gamma $, where $\gamma \sim 0.22$eV is the coupling constant upon using $eE_z = 3.56$eV/A$^\circ$. 
Then, the interaction matrix in the RPA-level charge susceptibility is given by 
\be
\hat{C} =  \biggl( J(\cos q_x + \cos q_y) +  \lambda \sum_{\k} \tilde{g}(\k,\q)\biggl) \hat{I},                                                                 
\ee\
where 
$\lambda_{} = 2\gamma^2/\omega_{B_{1g}} $. The physical charge susceptibility is obtained by taking the trace of the susceptibility matrix given by Eq. (13).
\section{results and Discussion}
\subsection{Charge correlation in normal state}
\begin{figure}
   \centering
\includegraphics[width=1.0\linewidth]{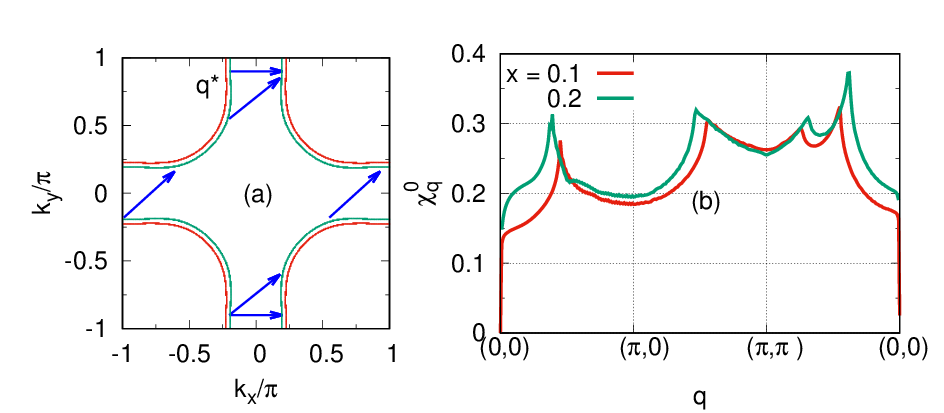}
   \caption{(a) Fermi surfaces for hole dopings $x = 0.1$ and $x = 0.2$. (b) Charge susceptibility $\chi_{\q}$ for $x = 0.1$ and $x = 0.2$ in the normal state without any long- or short-range order. }
    \label{f1}
    \end{figure}

We begin with the charge susceptibility calculation in the normal state for the hole dopings $x = 0.1$ and $0.2$ lying in the pseudogap region of the hole-doped cuprates (Fig.~\ref{f1}(a)). The neck-like sections of the Fermi surfaces are nearly straight, which leads to a good nesting between these sections. The distance between two nearly parallel sections decreases with an increase in doping as expected for the holes pockets. In particular, a good nesting associated with three wavevectors ($q^*,0$), ($\pi, \pi-q^*$) and ($q^*, q^*$) are noted. The same is confirmed by the peaks location of the charge susceptibility. The susceptibility, however, shows the largest peak at the bidirectional wavevector ($q^*, q^*$) instead and not at other wavevectors including the unidirectional one ($q^*,0$) observed experimentally (Fig.~\ref{f1}(b)). The origin of the largest peak at ($q^*, q^*$) is not surprising because the diagonal vector is able to connect all four pair of nearly straight and parallel running neck-like sections of the Fermi surfaces whereas the unidirectional wavevector ($q^*,0$) can connect only two pairs of neck-like sections. 
\begin{figure}
\centering
   \includegraphics[width=1.0\linewidth]{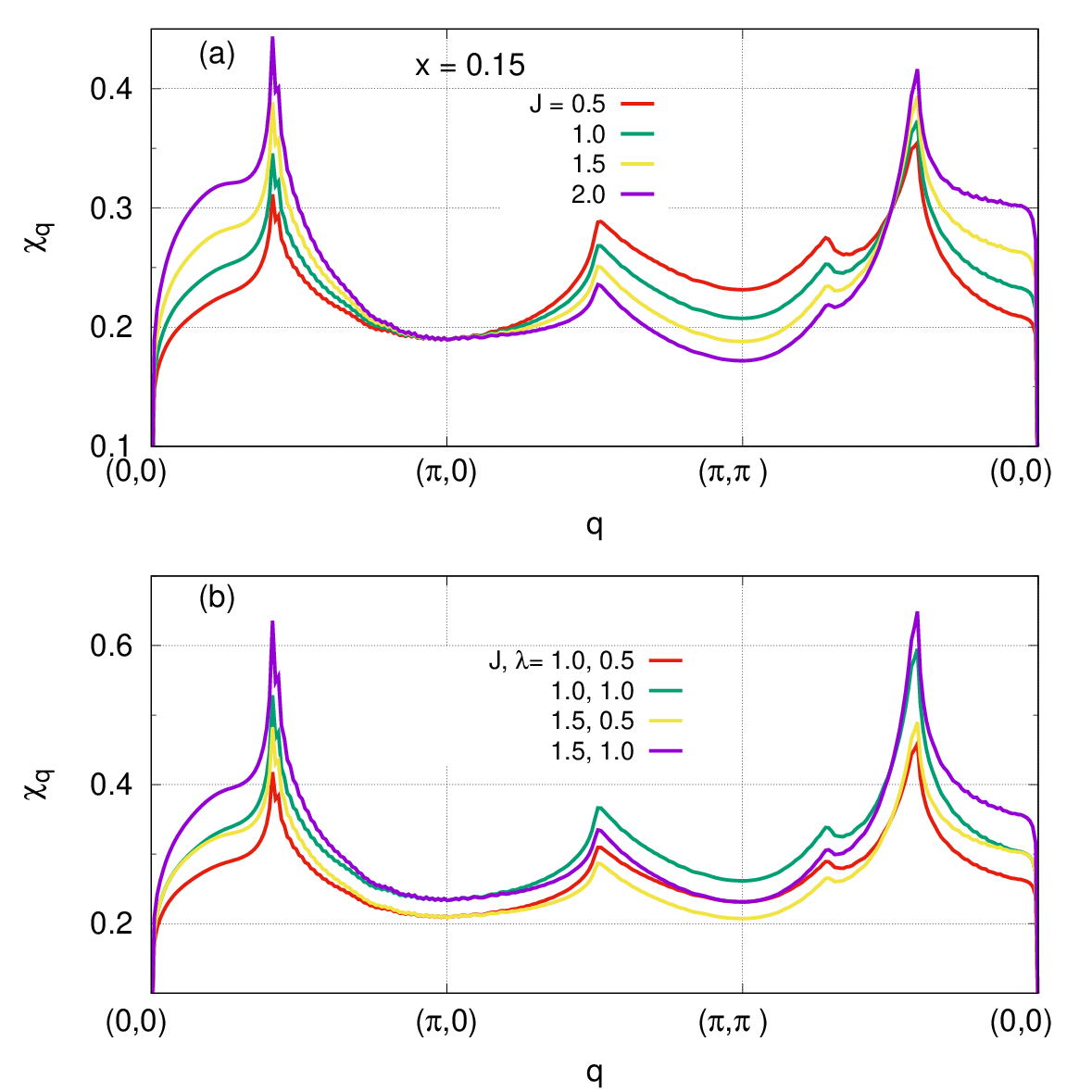}
   \caption{RPA-level susceptibility for hole doping $x = 0.15$ in the normal state. The charge susceptibility in the $t$-$J$ model (a) without and (b) with electron coupled to dispersionless phonon.}
    \label{f2}
    \end{figure}
    \begin{figure}
   \includegraphics[scale=1.0, width=9cm]{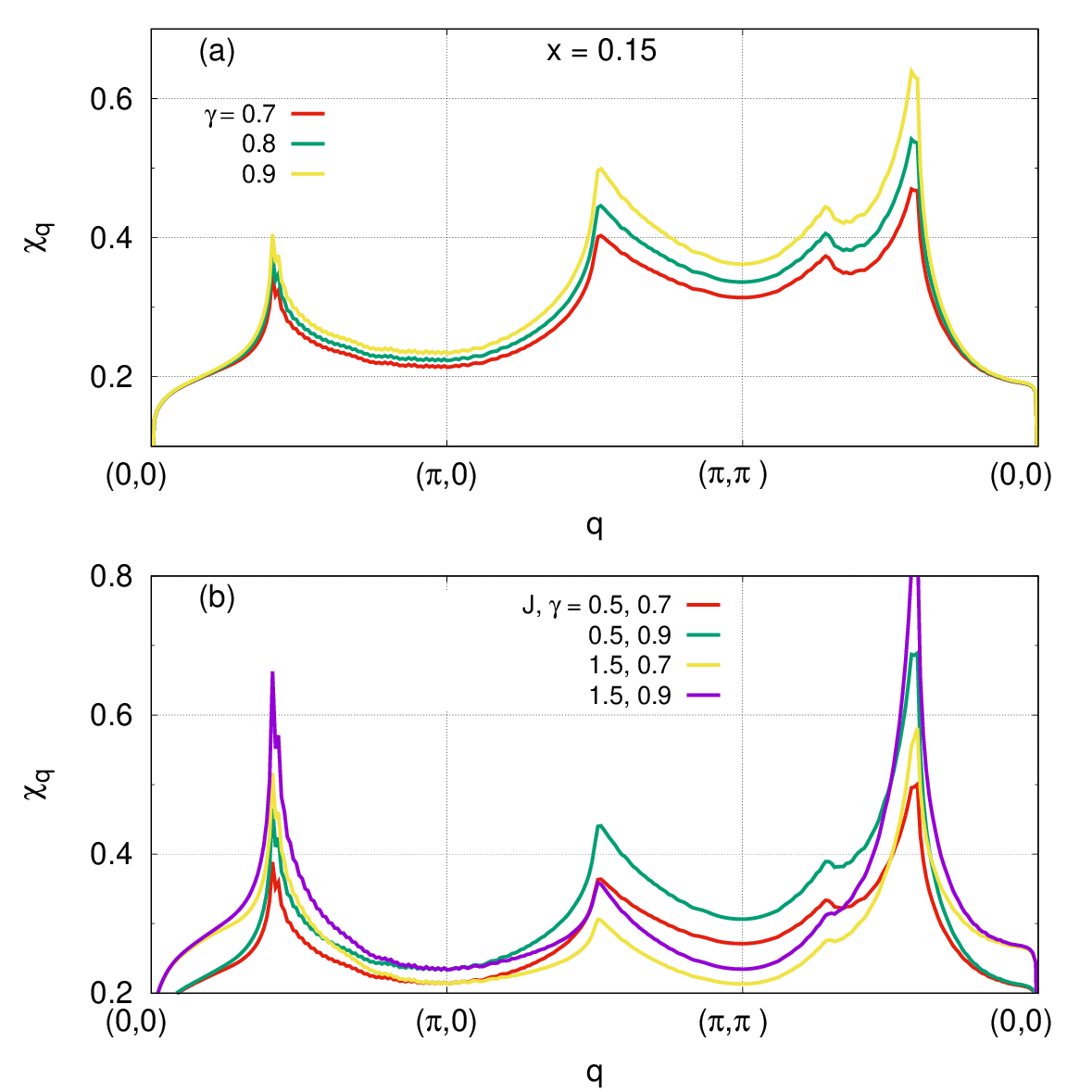}
   \caption{RPA-level susceptibility for hole doping $x = 0.15$ in the normal state. (a) The charge susceptibility as a function of electron-phonon coupling strength in the presence of only $B_{1g}$ phonons, where electron-phonon coupling $\gamma = \sqrt{\frac{\lambda \omega_{B_{1g}}}{2}}$. (b) The charge susceptibility in the $t$-$J$ model when coupling of electron to $B_{1g}$ phonons is taken into account.  }
    \label{f3}
    \end{figure} 
    \begin{figure}
   \includegraphics[scale=1.0, width=9cm]{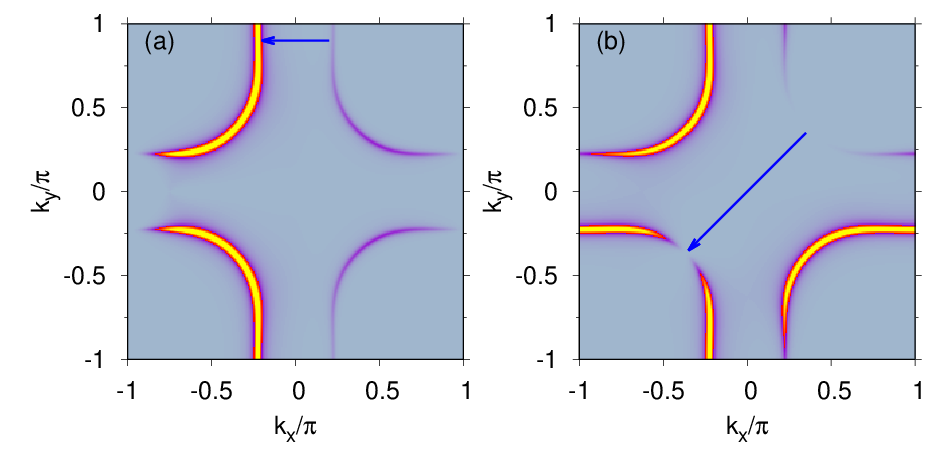}
   \caption{$|g(\k, \q)|^2$ plotted for initial momenta $\k^*$ at the Fermi surface  near (a) the antinodal point and (b) the nodal point while the final momenta $\k^{\prime} = \k^* + \q $ also on the Fermi surface. The arrows show final quasiparticle momenta, which is enhanced after getting scattered by the $B_{1g}$ phonon. }
    \label{f4}
    \end{figure}    
\begin{figure}
   \includegraphics[scale=1.0, width=9cm]{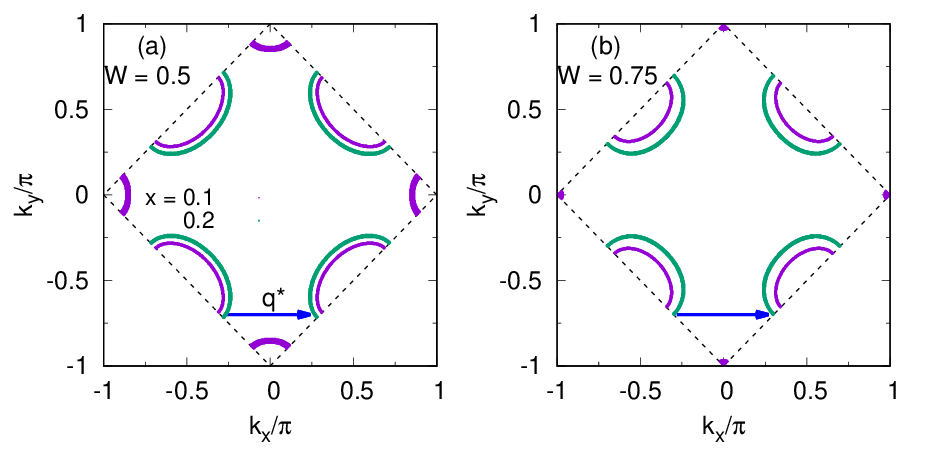}
   \caption{(a) Fermi surfaces for two hole dopings $x$ = 0.1 and 0.2 in DDW state for the two cases of interaction strength $W = 0.5$ and 0.75.}
    \label{f5}
    \end{figure}    
\begin{figure}
   \includegraphics[scale=1.0, width=9cm]{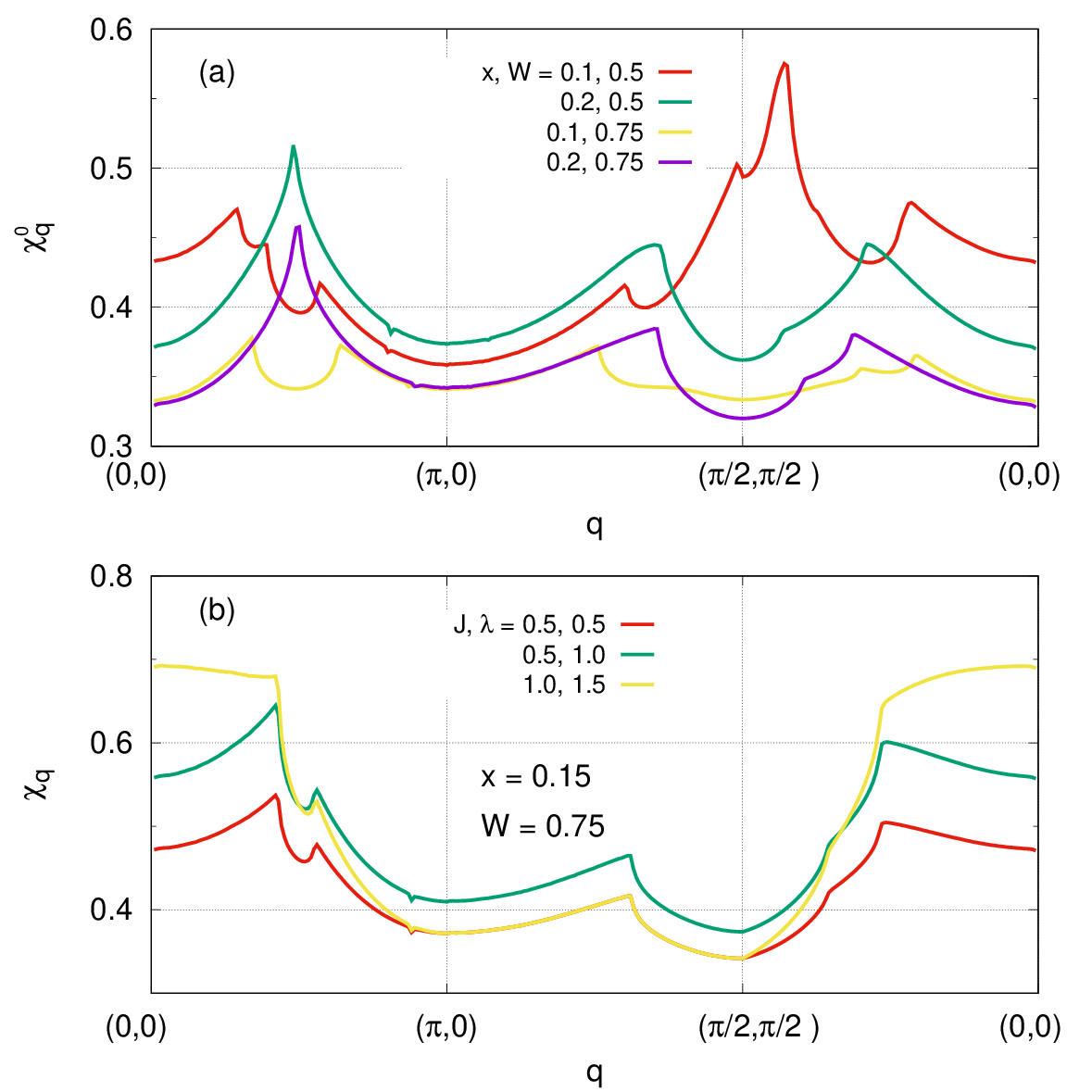}
   \caption{Charge susceptibilities in the state with DDW ordering. (a) Susceptibility for different dopings and interaction strengths. (b) RPA-level charge susceptibility in the $t-J$ model with electron-phonon coupling, when the phonon being dispersionless.}
    \label{f6}
    \end{figure} 
    \begin{figure}
   \includegraphics[scale=1.0, width=9cm]{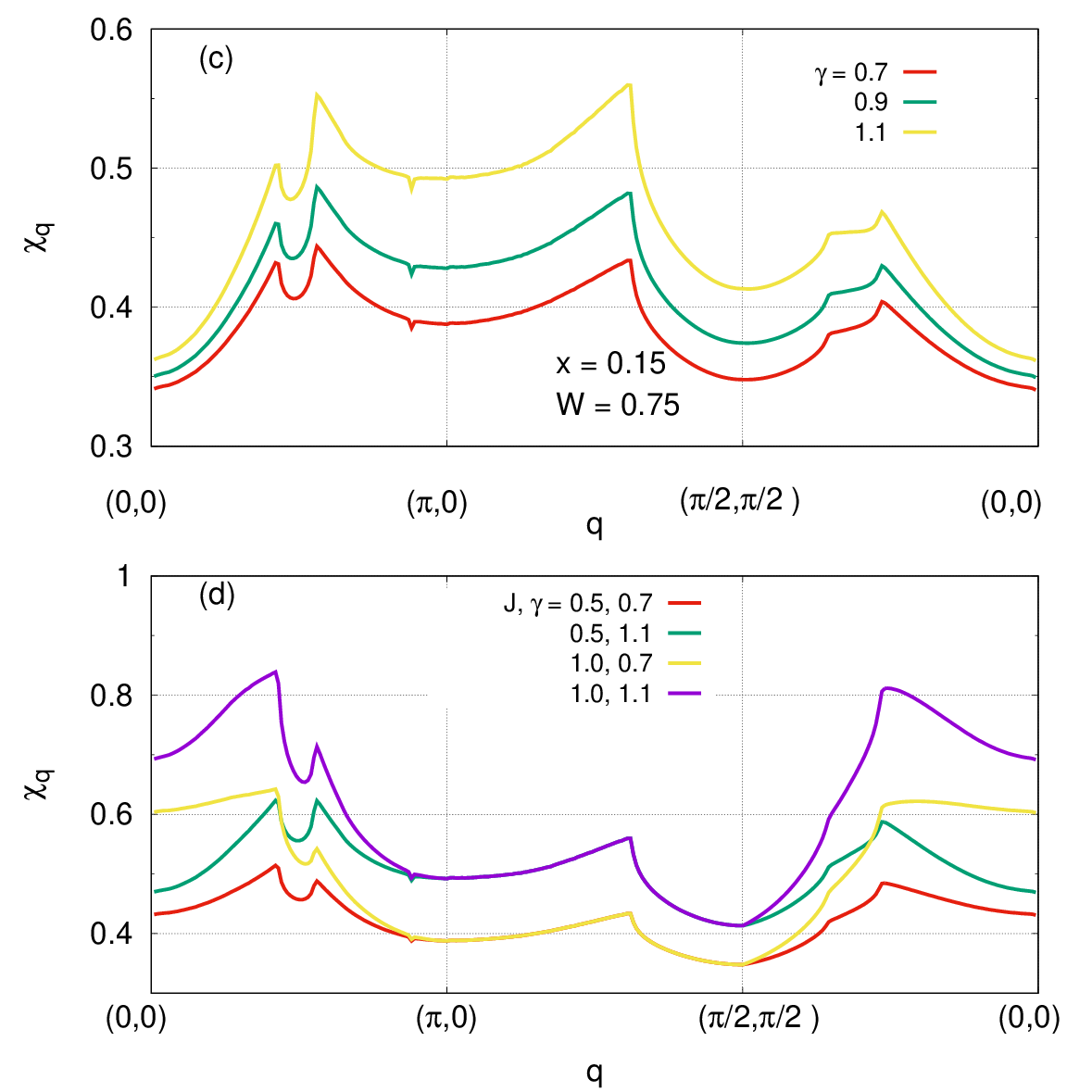}
   \caption{RPA-level charge susceptibilities in $d$-density wave state. (a) RPA-level charge susceptibility in the presence of only $B_{1g}$ phonons. (b) RPA-level susceptibility in the presence of nearest-neighbor Coulomb interaction and $B_{1g}$ phonons.}
    \label{f7}
    \end{figure} 
    
     \begin{figure}
   \includegraphics[scale=1.0, width=9cm]{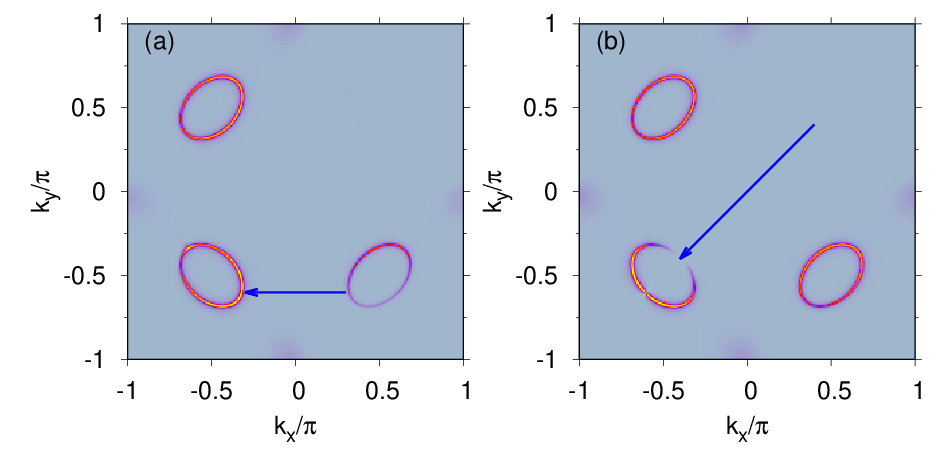}
   \caption{In the DDW ordered state, $|g(\k, \q)|^2$ plotted for initial momenta at the Fermi surface around the nodal points while the final momenta oriented along (a) $k^*_x$ and (b) diagonal direction while the final momenta $\k^{\prime} = \k^* + \q $ also on the Fermi surface. The arrows show final quasiparticle momenta, which is enhanced for unidirectional momentum and suppressed for bidirectional momentum after getting scattered by the $B_{1g}$ phonon.}
    \label{f8}
    \end{figure}
Fig.~\ref{f2} shows the RPA-level charge susceptibility calculated for the normal state in the presence of attractive long-range Coulomb interaction and electron-phonon coupling. First, we look at the susceptibility when only the long-range Coulomb interaction is included. In that case, the susceptibility, for a small interaction parameter $J$, exhibits peaks at diagonal and unidirectional wavevector of similar height while the peak near ($\pi, \pi-q^*_x$) is relatively suppressed. This feature can be understood from the momentum-dependent form of interaction $\propto \cos q_x + \cos q_y$. The magnitude of interaction is very small near $(\pi/2, \pi/2)$ as well near $(\pi, 0)$ while being the largest near (0, 0). Note that the interaction near $(\pi, \pi)$ will be negative and therefore the charge susceptibility is expected to be suppressed. It may be noted that the peak for the unidirectional wavevector becomes dominant only for unrealistically large $J = 2t^2_1/U \gtrsim 1$ as shown in Fig.~\ref{f2}(a). Inclusion of the dispersionless phonon has almost no effect except overall and nearly uniform increase in the susceptibility so that the peak for bidirectional wavevector dominates [Fig.~\ref{f2}(b)]. 

\subsection{Role of bond-buckling phonon and charge correlation in normal state}

On the other hand, the coupling of electron to the bond-buckling $B_{1g}$ phonon results in the enhancement of all the peaks [Fig.~\ref{f3}(a)] but the largest enhancement is for the incommensurate wavevector near $(\pi/2, \pi/2)$ in contrast with the experiments. This follows from the behavior of $|g(\k, \q)|^2$ plotted [Fig.~\ref{f4}] for initial unidirectional quasiparticle momentum pointing along the positive $k_x$ in the upper neck of the Fermi surface, while the final momenta $\k^{\prime} = \k + \q $ also lies on the Fermi surface. The arrow shows the final quasiparticle momenta, for which  scattering is enhanced by the $B_{1g}$ phonon [Fig.~\ref{f4}(a)]. A similar result is obtained for the bidirectional wavevector. However, diagonal scattering is suppressed if the initial momentum corresponds to a nodal point [Fig.~\ref{f4}(b)]. Fig.~\ref{f3}(b) shows the susceptibility in the $t$-$J$ model when the electronic interaction with $B_{1g}$ phonon is present. The RPA-level charge susceptibility continues to be dominant at a bidirectional wavevector. 
Thus, it appears that the normal-state Fermi surface cannot support the charge correlation being dominant one at the unidirectional wavevector. Next, we examine the charge susceptibility in the DDW state.

\subsection{Charge correlation in DDW state}
The DDW state, which gives rise to  weak magnetic moment
arising due to the circulating bond currents, can gap the Fermi surface $(\pi, 0)$ in a way similar to what is observed experimentally. As shown in Fig.~\ref{f5}, there are elliptical hole pockets only at the nodal points. The pocket size is sensitive to the interaction parameter $W$. Electron pockets can appear near ($\pi, 0$) for smaller hole doping and interaction parameter. In addition, the sensitivity of the hole or electron pockets to the hole doping in the ordered state may also be noticed especially when it is compared with normal state. Thus, the Fermi surfaces obtained in the DDW state qualitatively describes several features of the pseudogap phase and therefore can be a reasonable basis to provide an insight into the origin of the charge correlations. 

Fig. \ref{f6}(a) shows the charge susceptibility in the DDW state. The susceptibility is very sensitive to the interaction parameter $W$ as well as to the hole doping $x$. For $x = 0.1$ and $W = 0.5$, three major peaks can be seen along the high symmetry directions. They are located near ~($q^*,0$)), ($\pi/4,\pi/4$)) and at another point slightly away from ($\pi/2,\pi/2$)) along ($\pi/2,\pi/2$)) $\rightarrow$ (0, 0). The most dominant peak is the one near ($\pi/2,\pi/2$)) arising due to nesting between the hole pocket and electron pockets near ($\pi, 0$). However, the peak near ($\pi/2,\pi/2$)) gets suppressed for a larger hole doping $x = 0.2$ as the electron pocket disappear. On the other hand, when the interaction parameter is increased to $W = 0.75$ or hole is doped so that the electron pocket disappears, the susceptibility peak at unidirectional wavevector ~($q^*,0$)) becomes the highest one. Thus, the origin of the unidirectional charge correlation can be linked to the hole pockets connected by unidirectional nesting vector.

Fig. \ref{f6}(b) shows the RPA-level susceptibility calculated in the DDW state in $t$-$J$ model with the dispersionless phonon. We find that, for a smaller $J = 0.5$, the dominant peak occur at ~($q^*,0$) irrespective of the electron phonon coupling. However, as $J$ approaches unity, which is a reasonable value of the parameter, the peak shifts to (0, 0). Thus, the nearest-neighbor attractive interaction or coupling of electron to dispersionless phonon may not robustly be able to explain the origin of charge correlation observed in the cuprates. 

\subsection{bond-buckling phonon and charge correlation in DDW state}
Fig. \ref{f7}(a) shows the RPA-level charge susceptibility while considering the coupling of electron only to the $B_{1g}$ phonon. Interestingly, we find that the dominating charge susceptibility peak occurs at the unidirectional wavevector ~($q^*,0$), which is nearly robust with respect to any change in the coupling strength. However, there is another wavevector at  
($q^*,\pi/4$) which a comparable peak occurs. But this peak gets suppressed when a long-range Coulomb interaction from the $t$-$J$ model is incorporated [Fig.~\ref{f7}(b)].
As noted earlier, in the normal state, the modulus square of electron-phonon coupling $|g(\k, \q)|^2$ is peaked for the anti-nodal fermion momentum when the scattering vector $\q = 2\k_F$ is unidirectional and connects the parts of neck-like section of the Fermi surface. In the DDW state, there are portions of the Fermi surfaces, which are located not far from the antinodal point, which, therefore, facilitates in enhancing the susceptibility for the unidirectional vector and not for the diagonal vector [Fig.~\ref{f8}]. Although, the increase in peak size of all the peaks of susceptibility with electron-phonon coupling is nearly same. This features in the presence of bond-buckling phonon appears to be further robust when the long-range interaction is included. As noted previously, for a reasonable $J \sim 1$, RPA-level charge susceptibility exhibits a peak near (0, 0), but that issue is absent when the bond-buckling phonon is incorporated. Thus, it appears that the bond-buckling $B_{1g}$ phonon is expected to play a crucial role in the setting up of charge correlations in the cuprates with unidirectional wavevector. 

\section{conclusions} 
To conclude, we investigated charge correlations in the pseudogap state modeled by the $d$-density wave order. The latter does not require any accompanied large magnetic moments like the unrealistic antiferromagnetic state with short- or long-range magnetic order as considered in earlier works. Nature of the Fermi surfaces obtained in the $d$-density wave ordered state shows good qualitative agreement with the experimental features including the small hole pockets near ($\pi/2$, $\pi/2$) while gapped Fermi surface in the vicinity of ($\pi$, 0). Therefore, the $d$-density wave ordering provides a good starting point for the study of charge correlations in the pseudogap phase. Our analysis of static charge susceptibility indicates that the charge correlation in the normal state dominates at the  bidirectional wavevectors for a realistic value of interaction. On the contrary, these correlations dominate at  unidirectional wavevectors in the $d$-density wave ordered state. In particular, we show that a crucial role is played by the interplay between the electronic bandstructure and the bond-buckling $B_{1g}$ phonon, and the latter provides robustness to the dominance of unidirectional wavevector. These unidirectional charge correlations may also lead to the phonon softening observed in the $d$-wave superconducting state which may coexist with the $d$-wave density ordering. 
\section*{acknowledgement}
D.K.S. was supported through
DST/NSM/R\&D HPC Applications/2021/14 funded by
DST-NSM and start-up research grant SRG/2020/002144
funded by DST-SERB.

\end{document}